\documentclass[12pt,letterpaper]{article}

\usepackage{setspace}

\usepackage{graphicx}

\usepackage{siunitx}

\usepackage[numbers,sort&compress,merge]{natbib}
\bibpunct[, ]{[}{]}{,}{n}{,}{,}
\makeatletter
\renewcommand\@biblabel[1]{(#1)}
\makeatother

\usepackage{amsfonts}
\usepackage{amsmath}
\usepackage{amssymb}
\usepackage{xcolor}
\newcommand{\n}{\nonumber \\}

\newcommand{\tns}[1]{\textbf{\textit{#1}}}
\newcommand{\tnsgrk}[1]{{\boldsymbol #1}}
\newcommand{\opr}[1]{{\hat {#1}}}
\newcommand{\bra}[1]{\langle{#1}|}
\newcommand{\ket}[1]{|{#1}\rangle}
\newcommand{\prj}[1]{\langle{#1}}
\newcommand{\ie}{\textit{i.e.},~}
\newcommand{\eg}{\textit{e.g.},~}
\newcommand{\etc}{\textit{etc.}}
\newcommand{\secref}[1]{section \ref{#1}}
\newcommand{\figref}[1]{Fig.~\ref{#1}}
\newcommand{\eqnref}[1]{eq.~(\ref{#1})}

\newcommand{\Eqnref}[1]{Eq.~(\ref{#1})}

\newcommand{\eqnsref}[2]{eqs.~(\ref{#1}) and (\ref{#2})}

\newcommand{\appref}[1]{appendix \ref{#1}}
\definecolor{mygreen}{RGB}{0,188,0}
\newcommand{\new}[1]{#1}
\newcommand{\reworked}[1]{#1}

\begin{document}

\title{Systematically Improvable Excitonic Hamiltonians for Electronic Structure Theory}
\author{
  Anthony D.~Dutoi$^{*}$,
  Yuhong Liu \\
  {\footnotesize \textit{Department of Chemistry, University of the Pacific, Stockton, California 95211, USA}} \\
  $^*$\texttt{\small adutoi@pacific.edu}
}
\maketitle

\begin{abstract}
We show here that the Hamiltonian for an electronic system may be written exactly
 in terms of fluctuation operators that transition constituent fragments between internally correlated states,
 accounting rigorously for inter-fragment electron exchange and charge transfer.
Familiar electronic structure approaches can be applied to the renormalized Hamiltonian.
For efficiency, the basis for each fragment can be truncated,
 removing high-energy local arrangements of electrons from consideration,
 and effectively defining collective coordinates for the fragments.
For a large number of problems
\new{
 (especially for non-covalently interacting fragments),
}
 this has the potential to fold the majority
 of electron correlation into the effective Hamiltonian,
 and it should provide a robust approach to incorporating difficult electronic structure problems into large systems.
The number of terms in the exactly transformed Hamiltonian formally scales
 quartically with system size, but this can be reduced to quadratic in the mesoscopic regime, to within an arbitrary error tolerance.
Finally, all but a linear-scaling number of these terms may be efficiently decomposed in terms of electrostatic
 interactions between a linear-scaling number of pre-computed transition densities.
In a companion article, this formalism is applied to an excitonic variant of coupled-cluster theory.
\end{abstract}

		\section{Introduction}

A persistent pursuit of the electronic structure community is the elimination of unnatural barriers to efficient, detailed simulation.  
Formally, the amount of information needed to fully specify the quantum state of an electronic system grows exponentially
 with the system size.
But this is juxtaposed against the common intuition that makes possible the chemical language of atoms,
 molecules, and functional groups, which is that these entities do not forgo their individual properties upon interaction
 with others, though they may be heavily perturbed.
The most straightforward approach to eliminating the unnecessary bulk of an exponentially large state space is to divide a 
super-system into sub-systems,
 or fragments 
\reworked{
 \cite{Gordon:2011:FMOReview,Richard:2012:FragUnifiedView,Collins:2015:EnergyBasedFragMethods,Raghavachari:2015:FragmentReview,Huang:2008:ElecStructSolids,Wesolowski:2015:EmbeddingReview}
}.

\reworked{
One shortcoming of presently available fragment-based electronic structure methods
 is that they either do not capture inter-fragment electron correlation at all
 or this is treated at the level of individual electrons.
}
For inter-fragment correlation, a reliance on integrals describing interactions between individual electrons
 will render any high-order correlation scheme inefficient because 
 the structures of accompanying local relaxations are essentially recomputed for each separate interaction between
 the electrostatically efficacious single-electron fluctuations.
\new{
High-order (beyond perturbative) treatments of inter-fragment electron correlation are desirable, however, in order to capture
 many-body interactions between fragments
 \cite{Cui:2006:InductionFourthOrder,Lao:2016:LargeSystemMBE,Podeszwa:2007:SAPTDFT3Body,Ambrosetti:2016:WavelikeVdW}
 (a ``body'' is a fragment).
In turn, even low-order treatments of inter-fragment electron correlation are sensitive to the description of local correlation,
 given the sensitivity of polarizabilities to correlation level \cite{Riley:2010:IntermolForceDissect}.
This coupling between local and long-range correlation is missing in standard local correlation models
 \cite{
  Forner:1985:LocalCorrelation,
  Stoll:1992:LocalCorrDiamond,
  Saebo:1993:LocalCorrelation,
  Schutz:2001:DomainCCSD,
  Maslen:2005:MP4TRIM,
  Subotnik:2006:SmoothLocalCCSD,
  Li:2010:ClusterInMolecule,
  Hattig:2012:PNOMollerPlesset,
  Kristensen:2012:DivideExpandConsolidate,
  Liakos:2015:PNOCClimits},
 to the detriment of the long-range dispersion that holds together large systems.
}

An unexplored approach to fragment-based electronic structure calculations is to attempt to rewrite
 the super-system Hamiltonian exactly in terms of interacting fluctuations among \textit{internally correlated}
 states of fragments,
 such that its form mimics the familiar field-operator expression for the \textit{ab initio} Hamiltonian.
The degrees
 of freedom in the global computation are then fluctuations of entire subunits.
By making truncations in this context,
 high-energy local arrangements of electrons are expunged from consideration, and
 an effective suppression of individual degrees of freedom results.
Given the relative complexity of intra-fragment electron correlation
 (as compared to the simple picture of inter-fragment interactions),
 it is reasonable to view the internal coordinates of fragments in terms of such collective motions.

For the sake of being concrete, let us assert that a super-system
 Hamiltonian $\opr{\mathcal{H}}$ may be expressed in the following manner
 \begin{eqnarray}\label{Hform}
  \opr{\mathcal{H}} = \sum_m\sum_{i_m,j_m} H^{i_m}_{j_m} ~ \opr{\tau}^{j_m}_{i_m}
                    ~~ + \sum_{m_1<m_2}\sum_{\substack{i_{m_1},j_{m_1} \\ i_{m_2},j_{m_2}}}
                         H^{i_{m_1} i_{m_2}}_{j_{m_1} j_{m_2}} ~ \opr{\tau}^{j_{m_1}}_{i_{m_1}} \, \opr{\tau}^{j_{m_2}}_{i_{m_2}}
                    ~~ + ~ \cdots
 \end{eqnarray}
The two-index fluctuation operators are analogues of paired field operators;
 the effect of $\opr{\tau}^{j_m}_{i_m}$ is to induce a fluctuation of fragment $m$ only,
 from some state $\ket{\psi_{j_m}}$ to another state $\ket{\psi_{i_m}}$.
(The positioning of indices is addressed later.)
This implies a basis of super-system states, denoted
 \begin{eqnarray}
  \ket{\Psi_I} = \ket{\psi_{i_1} \psi_{i_2} \cdots }
 \end{eqnarray}
 which have tensor-product-like structure in terms of internally correlated fragment states $\ket{\psi_{i_m}}$.
The collection of fragment-state labels into a single ket implies global antisymmetry of the electronic wavefunction.
The elements $H^{i_m}_{j_m}$ in \eqnref{Hform} build a Hamiltonian matrix for fragment $m$, and the higher-order terms are responsible for
 couplings between fragments ($m_1$$<$$m_2$ under the summation runs over all unique pairs).

\begin{figure}
  \centering
  \includegraphics[width=14.5cm]{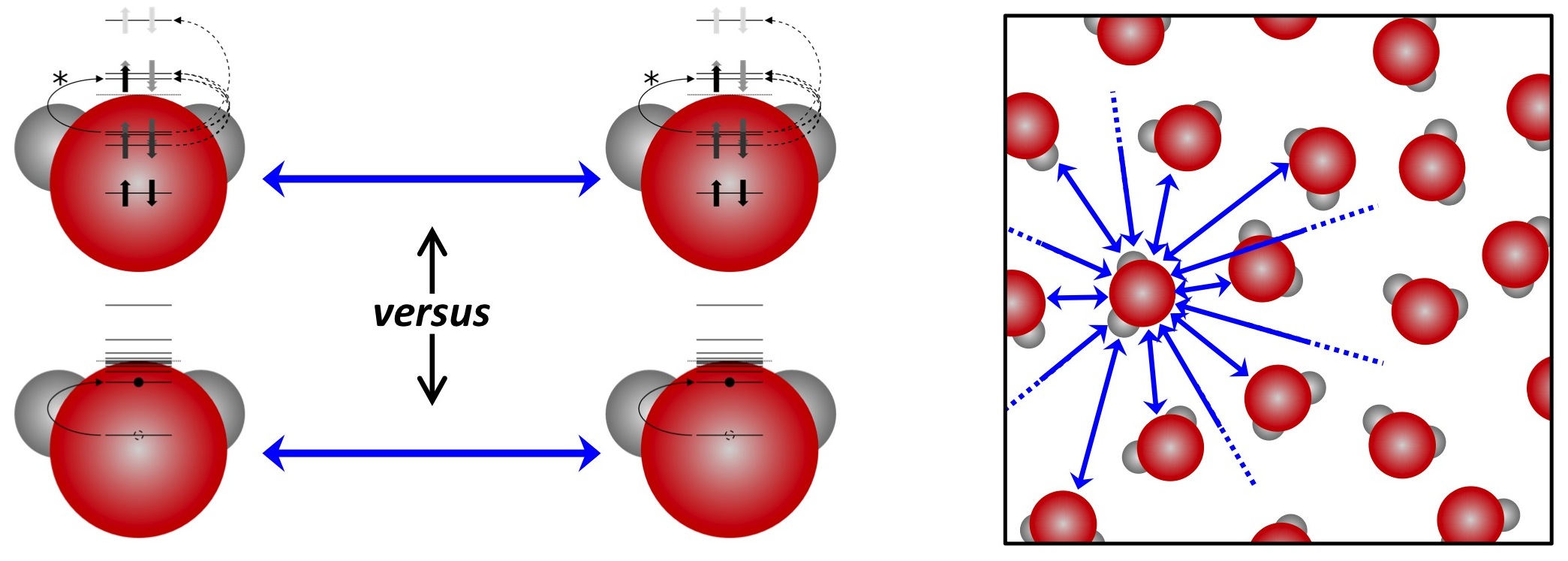}
  \caption{\label{concept}
    An interaction between a pair of molecules can be thought of in terms of
     individual electrons fluctuating among local orbitals, whereby a ``primary''
     excitation (denoted with an asterisk) is accompanied by a number of other connected relaxations
     (nominal excitations).
    The same interaction can also be conceived of in terms of fluctuations
     between electronically correlated states of the fragments, which inherently
     contain these relaxations.
    Given the relatively large energetic scale of intra-fragment correlation,
     the relaxations that
     accompany local fluctuations should be
     largely similar across interactions
     \new{
     of one fragment with many neighbors, as illustrated in the panel to the right.
     As such, the structure of these relaxations
     }
     need only be computed once for each fragment.
  }
\end{figure}

The internally correlated fragment states will be referred to as the excitonic basis,
 since fluctuations among these
 are the conceptual site basis from which Frenkel--Davydov
 excitons \cite{Frenkel:1931:Excitons,Davydov:1971:MolecularExcitons,May:2011:ChgNrgXfer}
 are built.
If the Hamiltonian can indeed be represented exactly in terms of such renormalized
 fluctuations, then recourse to the primitive electronic degrees of freedom can be avoided at the global scope.
The intuition behind the expected efficiency of the proposed scheme is illustrated pictorially in \figref{concept}.
Local relaxations that accompany the primary motions of fluctuating electrons are computed once for each fragment,
 and these are permanently folded into the effective Hamiltonian.
As the fragment bases are truncated to reduce the dimensionality of the global calculation,
 strong intra-fragment correlations remain constant features of the retained state space that describes low-energy phenomena.
Solving for intra-fragment correlation first and then interacting the fragments in the transformed picture
 essentially inverts the traditional paradigm of using a fragment-based decomposition
 of a reference wavefunction as a starting point for handling global electron
 correlation.

Traditional electronic structure approaches
 are available to treat inter-fragment interactions in the excitonic picture.
\reworked{
 In principle, even the covalent bonding interaction can be represented by allowing correlated charge-state fluctuations,
}
\new{
 though excitonic renormalization is probably most useful for weakly interacting fragments.
}
System-wide induction can be captured at the mean-field level, but using polarizabilities
 from internally correlated fragments.
The usual post-mean-field approaches (\eg{perturbation theory, coupled-cluster theory}),
 can be used to handle inter-fragment electron correlation (\ie{Van der Waals forces}).
\new{
Excitonically renormalized Hamiltonians might also be used in dynamical mean-field theory
 calculations for the electronic structures of crystals \cite{Kotliar:2006:DynamicalMeanField}.
}
The computational cost for a global calculation (after the effective Hamiltonian is obtained)
 will depend only on the numbers of fragments and states per fragment,
 not on the internal structure of those fragment states.

\reworked{
Block correlated (BC) methods \cite{Shen:2009:BCCCexcitedstates,Xu:2013:BCPT2}
}
\new{
 and the active-space decomposition (ASD) approach \cite{Parker:2014:ActiveSpaceDecompDMRG,Kim:2015:ActiveSpaceCovalent} both
}
\reworked{
 use a partitioning of electron correlation that is formally similar to this work.
However, BC methods were not conceived as fragment methods (rather used to isolate strong correlations in a molecule),
}
\new{
 and work on the ASD approach has not yet allowed for charge transfer or developed a field-operator-like Hamiltonian resolution that would be amenable to, say, coupled-cluster theory.
Importantly, both BC and ASD were also only proposed for non-overlapping (explicitly orthogonalized) subsystems,
 such that recovering the use of the full basis assigned to any fragment would involve formal charge transfers to its neighbors.
}
The concept of working with general fragment states in an \textit{ab initio} method is also not generally new.
Most notably, it is prominent in the formal development of symmetry-adapted perturbation theory \cite{Jeziorski:1994:SAPT,Hohenstein:2010:SAPTtriples}.
However, the computational implementation still proceeds in terms of a Hamiltonian described in the one-electron basis,
 and extension of the globally antisymmetrized working equations beyond dimers is algebraically tedious.
Other recent and compelling work on \textit{ab initio} exciton theory \cite{Sisto:2014:AbInitoExciton,Morrison:2015:AbInitioExciton}
 has focused specifically on the excited-state regime;
 however, these approaches currently lack a clear and tractable scheme for systematic inclusion of ever higher levels of electron correlation,
 which the approach should provide.

The purpose of this article is to demonstrate rigorously that the Hamiltonian for a
 fragment-decomposed electronic system may indeed be written exactly in the form asserted in \eqnref{Hform},
 also when fragment orbitals overlap each other.
We furthermore provide practicable recipes for building such excitonic Hamiltonian resolutions.
The difficultly in developing rigorous and computationally tractable definitions for the super-system basis and associated fluctuation operators
 for overlapping fragments
 lies with inter-fragment electron-exchange antisymmetry.
We start by next giving a conceptual outline of the loosely deductive
 process that leads to the resolution of the exchange problem pursued here.
The bulk of the article is then spent making those concepts rigorous, in
 a bottom-up procedural manner.
A final section before the conclusion is dedicated to
 discussing the practical aspects of computing the necessary matrix elements,
 including the applicability of approximation schemes.

\section{Approach to the Inter-fragment Exchange Problem \label{exchange_concept}}

The root of the exchange problem
\new{
 (\ie{the practical difficulty of enforcing wavefunction antisymmetry})
}
 is that
 the one-electron spaces describing each fragment cannot
 be orthogonalized against each other without damaging the descriptions of the 
 correlations internal to the fragments
 (and in ill-defined ways that would depend on the orthogonalization scheme).
We therefore take as a premise that the one-electron spaces available to each of the fragments are not to be redefined.
Due to Pauli exclusion then,
 the available space on one fragment will be dynamically entangled with the fluctuating,
 internally correlated electronic structures of its neighbors.
The fact that a fluctuation on one fragment then effectively changes the states of all of its neighbors
 initially casts doubt on whether the Hamiltonian can be expressed
 in terms of a \textit{low-order} expansion of single-fragment fluctuation operators.

From a more formal perspective,
 although it is trivial to assert a definition of an antisymmetrized tensor product of internally correlated fragment states $\ket{\Psi_I}$,
 a basis of such super-system states is not generally orthonormal,
 a reflection of the overlap-driven exchange interaction.
The lack of an orthonormal basis is
 naturally handled by the introduction of biorthogonal complements, denoted as $\ket{\Psi^I}$.
The unique association of each complement with precisely one of the original tensor-product states
 can be leveraged to construct operators effecting single-fragment fluctuations between the original basis states.
Although this step is relatively straightforward, the matrix elements of the consequent
 Hamiltonian expansion depend critically on the nature of the states used to construct them.
This moves the apparent difficulty of the electron-exchange problem
 into the construction of suitable definitions of the complements.

The biorthogonal complements are, in principle, described
 by the inverse of the matrix of overlaps $\prj{\Psi_I}\ket{\Psi_J}$.
This is purely formal, however,
 since direct numerical inversion of such an exponentially large matrix is intractable,
 especially since it will not be sparse nor exactly factorizable.
Let us consider more fundamental
 questions of structure, however,
 under the presumption that the biorthogonal basis is somehow available.
First, the complements $\ket{\Psi^I}$ will not themselves have tensor-product-like structure in terms of the original fragment states;
 they will appear to contain inter-fragment correlations.
Furthermore, 
 for a given \textit{truncation} of the fragment bases,
 a different effective Hamiltonian may be obtained,
 depending on how the complements are defined.
Such truncations are critical for the efficiency reasons outlined in the introduction.
The ambiguity in the definition of the complements results because
 the projected inverse of a matrix is not the same as the inverse of a projected matrix.
Insisting that the complements live in the post-truncation model space would be different
 than choosing them to be the corresponding members of the set of complements to 
 the untruncated tensor-product set, or any other set.
This unsettling ambiguity would
 likely be tolerable as an approximation for sufficiently large model spaces,
 but the choice of complements also affects
 the formal fragment order of the Hamiltonian expansion.

There is one particular choice of complements with simple structure, amenable to algebraic manipulation;
 these are the full-space complements of the complete tensor-product set
 of fragment bases (which themselves span the complete Fock spaces of their respective fragments).
Given a truncated model space, the complements therefore span a different subspace, 
 but of the same dimensionality.
The nature of their simplicity
 is that each complement does, in fact, have tensor-product-like structure,
 but in terms
 of secondary sets of states 
 associated with each fragment.
Each of the members of the implied 
 secondary fragment bases
 has unit overlap with exactly one of the original states of its associated fragment,
 and it is strictly orthogonal to all other fragment basis states,
 both on that same fragment or on any other fragment.
This is the key to achieving both the biorthogonality needed to define the fluctuation operators
 and the fragment locality needed obtain a low-order Hamiltonian expansion
\new{
 with tractable matrix elements.
}

The forthcoming procedure
 starts by systematically constructing
 definitions of biorthogonal one-electron, single-fragment, and super-system bases.
Once the super-system bases are defined, the relatively straightforward definition of the
 fluctuation operators in terms of them is presented.
Using the properties of the bases forwarded, the corresponding matrix elements in the excitonic Hamiltonian expansion
 may be deduced, showing that it has a low-fragment-order resolution.

		\section{The Excitonic Hamiltonian for Electronic Systems}

	\subsection{Notation and conventions}

The scope of this work is limited to the non-relativistic electronic Hamiltonian, projected into a basis,
 for fixed nuclear positions.
The system is divided into $N$ enumerated fragments, which are disjoint groups of atoms.
For conceptual ease, we presume that the atoms of each fragment occupy the spatial positions that they have
 in the super-system, but the many-electron state space associated with each
 is defined as though it were in isolation at that location.

Lower-case latin letters will be used for integer indices, and upper-case latin
 letters will be used for ascending-ordered tuples of integers,
 for example, $I = (i_1, \cdots i_{\ell_I})$, where
 $\ell_I$ is the length of tuple $I$.
The greek letters $\Psi$ and $\psi$ will be used to refer to general states of the super-system and fragments, respectively,
 and $\Phi$ and $\phi$ will refer to the respective single-determinant states from which these are built.
We use $\chi$ for one-electron orbitals.

On matrix-valued quantities, subscripts and superscripts will be used to index covariant and contravariant dimensions, respectively.
If a matrix has both a covariant and a contravariant index, then, for purposes of matrix multiplication,
 the rows are to be enumerated by the
 contravariant (typically bra) index.
Sets are abbreviated as $\{y_i\}$ to represent all $y_i$ corresponding to
 values of $i$ that are defined for mapping $y$.
Summations implicitly run over all values of an index that are allowed by the mapping to which the index is attached.

	\subsection{The one-electron bases}

We presume a set of linearly independent orbitals $\{\ket{\chi_p}\}$,
 where each orbital is associated with a specific fragment
 (\eg{atomic orbitals of constituent atoms, molecular orbitals of fragments, \etc}).
The index that enumerates this set is taken to be ``blocked'' by fragment,
 such that the first block of consecutive values
 enumerates the orbitals on fragment 1, and so forth.
For ease of discussion,
 we take the space spanned by the one-electron basis for a given problem
 as our working definition of complete,
 consequently defining 
 what is meant by completeness
 of a many-electron Hilbert space and completeness of the Fock space
 (having electron numbers up to the cardinality of $\{\ket{\chi_p}\}$).
%

We let the complete set of biorthogonal complement orbitals be denoted $\{\ket{\chi^p}\}$, such that
  $\prj{\chi^p}\ket{\chi_q} = \delta_{pq}$.
Should a linearly dependent set of functions be proposed, then some difficulty arises.
The resolution of this (discussed 
\reworked{
 in \appref{lin_alg})
}
 maps the linearly dependent case onto a problem of the same structure;
 therefore, the remainder of this discussion is general.

Field-operator notation will be convenient for developing transparent rules for matrix elements,
 and there are some important subtleties when working with biorthogonal bases
\reworked{
 \cite{Helgaker:2002:PurpleBook,HeadGordon:1998:Biorthogonal}.
}
Here we let $\opr{c}_p$ and $\opr{c}^p$ denote field operators that correspond to creation
 of $\ket{\chi_p}$ and $\ket{\chi^p}$, respectively, within a single-determinant state.
The Hermitian conjugates of these operators, denoted $\opr{a}_p$ and $\opr{a}^p$, respectively,
 act as annihilation operators of their \textit{complements} within ket states.
The biorthogonal field operators obey the following anticommutation relationships (among others)
 \begin{eqnarray}\label{bo_anticomm}
  \big[ \opr{c}_p , \opr{a}^q \big]_+ = \delta_{pq} \;\;\;\;\; \;\;\;\;\;
  \big[ \opr{c}_p , \opr{c}_q \big]_+ = 0 \;\;\;\;\; \;\;\;\;\;
  \big[ \opr{a}^p , \opr{a}^q \big]_+ = 0
 \end{eqnarray}
 allowing us to write
 the \textit{ab initio} Hamiltonian as
 \begin{eqnarray}\label{field_opH}
  &~& \opr{\mathcal{H}} = \sum_{p,q} h^p_q ~ \opr{c}_p \opr{a}^q + \sum_{p,q,r,s} v^{pq}_{rs} ~ \opr{c}_p\opr{c}_q\opr{a}^s\opr{a}^r \n
  &~& h^p_q = \bra{\chi^p} \opr{h} \ket{\chi_q} \;\;\;\;\;
  v^{pq}_{rs} = \frac{1}{4} \bra{\chi^p \chi^q}\opr{v}\ket{\chi_r \chi_s}
 \end{eqnarray}
 where $\opr{h}$ is the combined kinetic energy and nuclear attraction operator,
 and $\opr{v}$ is the electron--electron repulsion operator.
\new{
 The two-electron integrals here are implicitly antisymmetrized since the bra and ket are Slater determinants.
}
\reworked{
While both integral tensors here lack the usual symmetry with respect to permuting bra and ket indices,
}
 the anticommutation relationships for the field operators recover all of the rules
 that are familiar from the orthonormal case
 for making orbital substitutions
 and taking matrix elements,
 so long as bra states are expressed in the complement basis.

	\subsection{Many-electron bases}

\subsubsection{Super-system determinant bases}

The complete Fock space of the super-system is spanned by the set of single-determinant configurations $\{\ket{\Phi_P}\}$ where 
 \begin{eqnarray}\label{detdef1}
  \ket{\Phi_P} ~=~ \ket{\chi_{p_1} \cdots \chi_{p_{n_P}}} 
               ~=~ \opr{c}_{p_1} \cdots \opr{c}_{p_{n_P}} \ket{~}
 \end{eqnarray}
The index $P = (p_1, \cdots p_{n_P})$ can be any
 ordered tuple of any allowed length (up to the basis size),
 where $n_P$ is the number of electrons in configuration $\ket{\Phi_P}$,
 and  $\ket{~}$ is the true vacuum.
For illustrative purposes,
 we note that $\ket{\Phi_P}$ could also be written as
 \begin{eqnarray}\label{detdef2}
  \ket{\Phi_P} = \sqrt{n_P!} \, \opr{\mathcal{P}}_\text{A} \big[ \ket{\chi_{p_1}} \otimes \cdots \ket{\chi_{p_{n_P}}} \big]
 \end{eqnarray}
 where
 $\opr{\mathcal{P}}_\text{A}$
 is an orthogonal projector from the space of all raw (asymmetric) orbital-tensor-product states
 onto the subspace of all antisymmetric states (\ie{the complete Fock space}).
\new{
The $\sqrt{n_P!}$ prefactor of \eqnref{detdef2} is necessary so that the norm of this state matches that in
 \eqnref{detdef1}, which would be unity if the orbitals were orthonormal.
}

If the component indices of a tuple $P$ are chosen such that they all identify orbitals on a single fragment, then
 the use of the lower case in $\ket{\phi_P}$ will emphasize this.
The set $\{\ket{\phi_P}\}$, a strict subset of $\{\ket{\Phi_P}\}$,
 is then the set of all possible single-fragment determinant states on all possible fragments.
If we wish to indicate the fragment $m$ to which a given such determinant belongs, then the notation $\ket{\phi_{P_m}}$ is used;
 the subscript $m$ is thought of as placing a restriction on the possible values of the tuple.
A summation over an index $P_m$ would run over only those determinant states of fragment $m$.

A general tuple $P$ may be subdivided into the (potentially empty) tuples $P_1$ through $P_N$, each containing only the component orbital indices
 of $P$ that belong to the fragment indicated by the subscript.
(It is simply convenient that the indexing of the sub-tuples of $P$ is coincident with the convention
 used to indicate fragment-based restrictions.)
Since all tuples in this work are taken to be ordered, and since the indexing of the orbitals is blocked by fragment,
 $P$ is always reconstructed by simple concatenation of the sub-tuples.
This then allows us to write
 \begin{eqnarray}\label{antisymmpdt}
  \ket{\Phi_P} ~=~ \ket{\phi_{P_1} \cdots \phi_{P_N}}
               &=& \opr{\phi}_{P_1} \cdots \opr{\phi}_{P_N} \ket{~} \n
               &=& \sqrt{n_P! / ( n_{P_1}! \cdots n_{P_N}!)} \, \opr{\mathcal{P}}_\text{A} \big[\ket{\phi_{P_1}} \otimes \cdots \ket{\phi_{P_N}} \big]
 \end{eqnarray}
 where $\opr{\phi}_{P_m}$ collects (in ascending order) the creation operators of the orbitals in $P$ that belong to fragment $m$.
The second line of \eqnref{antisymmpdt} is shown to be equal to $\ket{\Phi_P}$ by careful manipulation of the nested antisymmetrizations
 as defined in \eqnref{detdef2},
 and this writing emphasizes the connection to a tensor product of fragment states.

In either notation for \eqnref{antisymmpdt}, it is clearly meaningful to 
 say that fragment $m$ is in state $\ket{\phi_{P_m}}$ in the globally antisymmetric super-system state $\ket{\Phi_P}$.
We have the intuitive result that the complete basis $\{\ket{\Phi_P}\}$
 is also described as the set of all antisymmetrized tensor products of determinants for the constituent fragments.

Using the same conventions as above, but applied to the orbitals $\{\ket{\chi^p}\}$,
 we have the set $\{\ket{\Phi^P}\}$, whose members satisfy
 \begin{eqnarray}\label{detcomplement}
  \bra{\Phi^P} ~=~ \bra{\phi^{P_1} \cdots \phi^{P_N}} &=& \bra{~} \opr{\phi}^{P_N} \cdots \opr{\phi}^{P_1} \n
               &=& \sqrt{n_P! / ( n_{P_1}! \cdots n_{P_N}!)} \, \big[\bra{\phi^{P_1}} \otimes \cdots \bra{\phi^{P_N}} \big] \opr{\mathcal{P}}_\text{A}
 \end{eqnarray}
 where the operator $\opr{\phi}^{P_m}$ is built from annihilation operators (with indices in descending order).
We can quickly verify that $\{\ket{\Phi^P}\}$ is the set of biorthogonal complements to $\{\ket{\Phi_P}\}$ by using
 the aforementioned anticommutation relationships
 \begin{eqnarray}
  \prj{\Phi^P}\ket{\Phi_Q}
    = \prj{\chi^{p_1} \cdots \chi^{p_n}}\ket{\chi_{q_1} \cdots \chi_{q_n}}
    = \big\langle \opr{a}^{p_{n_P}} \cdots \opr{a}^{p_1} \opr{c}_{q_1} \cdots \opr{c}_{q_{n_Q}} \big\rangle = \delta_{PQ}
 \end{eqnarray}
 where the angle brackets denote a vacuum expectation value.
This makes use of the fact that, since the tuples are ordered, if $P \neq Q$, then they are different in composition.

\subsubsection{General tensor-product super-system bases}

We can now define a basis for the super-system Fock space, in terms of more general fragment states.
Let the set $\{\ket{\psi_i}\}$ collect all such basis states for all fragments.
These are defined by introduction of an invertible matrix $\tns{z}$ with elements $z^P_i$, such that
 \begin{eqnarray}\label{general_frag}
  \ket{\psi_{i_m}} 
                   ~=~ \sum_{P_m} z^{P_m}_{i_m} \, \ket{\phi_{P_m}}
                   ~=~ \Big( \sum_{P_m} z^{P_m}_{i_m} ~ \opr{\phi}_{P_m} \Big) \ket{~}
                   ~=~ \opr{\psi}_{i_m} \ket{~}
 \end{eqnarray}
The square matrix $\tns{z}$ has rows indexed by a tuple and columns indexed by an integer.
The subscript on $i_m$ restricts the index
 to refer to one of the states of fragment $m$.
As an artifact of our notation for blocking indices by fragment,
 $\tns{z}$ formally has elements that would mix determinants on separate fragments, but we insist that it is block-diagonal by fragment;
 when necessary, we let the diagonal block of $\tns{z}$ for fragment $m$ be denoted as $\tns{z}^{(m)}$.
For simplicity, we will also presume that $\tns{z}$ does not mix determinants of different particle number,
 though some of the following could be generalized beyond this.

Using the biorthogonality of the fragment-determinant bases,
 it is straightforward to show that $\prj{\psi^i}\ket{\psi_j} = \delta_{ij}$ for the set $\{\ket{\psi^i}\}$, which satisfy
 \begin{eqnarray}\label{general_comp}
  \bra{\psi^{i_m}} 
                   ~=~ \sum_{P_m} \bar{z}^{i_m}_{P_m} \, \bra{\phi^{P_m}}
                   ~=~ \bra{~} \Big( \sum_{P_m} \bar{z}^{i_m}_{P_m} ~ \opr{\phi}^{P_m} \Big)
                   ~=~ \bra{~} \opr{\psi}^{i_m}
 \end{eqnarray}
 where, for convenience, we use the notation $\bar{\tns{z}} = \tns{z}^{-1}$, which has elements $\bar{z}^i_P$.
Since
 $\bar{\tns{z}}$ is also block-diagonal by fragment
 the notation $\ket{\psi^{i_m}}$ to refer to a state associated with fragment $m$ is logically sensical,
 regardless of ambiguous physical interpretation as such.

We now construct another pair of complete biorthogonal bases for the super-system Fock space, $\{\ket{\Psi_I}\}$ and $\{\ket{\Psi^I}\}$, where
 \begin{eqnarray}\label{fieldopstates}
  \ket{\Psi_I} ~=~ \ket{\psi_{i_1} \cdots \psi_{i_N}} &=& \opr{\psi}_{i_1} \cdots \opr{\psi}_{i_N} \ket{~} \n
               &=& \sqrt{n_I! / (n_{i_1}! \cdots n_{i_N}!)} \, \opr{\mathcal{P}}_\text{A} \big[\ket{\psi_{i_1}} \otimes \cdots \ket{\psi_{i_N}} \big] \n
  \bra{\Psi^I} ~=~ \bra{\psi^{i_1} \cdots \psi^{i_N}} &=& \bra{~} \opr{\psi}^{i_N} \cdots \opr{\psi}^{i_1} \n
               &=& \sqrt{n_I! / (n_{i_1}! \cdots n_{i_N}!)} \, \big[\bra{\psi^{i_1}} \otimes \cdots \bra{\psi^{i_N}} \big] \opr{\mathcal{P}}_\text{A}
 \end{eqnarray}
 where $I=(i_1,\cdots i_N)$ gives the indices of the states of each of the sub-systems.
The field-operator notation is an elegant manner to enforce global antisymmetry,
 whereas the tensor-product notation connects more directly to the state-space descriptions of the fragments.
It is notable that the state-space definition of antisymmetrization for general states is the
 same as that applied to single-determinant states;
 although it does not rely on permutation operators or determinant arithmetic,
 it does rely on the fragment states having definite particle number.
From either approach to antisymmetrization, the following connections between the two pairs of biorthogonal bases are obtained
 \begin{subequations}
 \begin{eqnarray}
  \ket{\Psi_I} &=& \sum_P Z^P_I \, \ket{\Phi_P} ~~;~~~~~~~~ Z^P_I = \prod_m z^{P_m}_{i_m} \label{superbasis} \\
  \bra{\Psi^I} &=& \sum_P \bar{Z}^I_P \, \bra{\Phi^P} ~~;~~~~~~~~ \bar{Z}^I_P = \prod_m \bar{z}^{i_m}_{P_m} \label{supercompl}
 \end{eqnarray}
 \end{subequations}
 where, in the definitions of the elements of $\tns{Z}$ and $\bar{\tns{Z}}$, the index $m$ runs over all of the fragments.
It is easy to verify that $\tns{Z}$ and $\bar{\tns{Z}}$ are mutually inverse, yielding $\prj{\Psi^I}\ket{\Psi_J} = \delta_{IJ}$.
Since $\tns{Z}$ is invertible, the set $\{\ket{\Psi_I}\}$ (and also $\{\ket{\Psi^I}\}$) is a complete basis.

In the most straightforward conceptualization, the orbitals on each
\new{
 individual
}
 fragment could be taken to
 be orthonormal among each other 
\new{
 (though overlapping the orbitals of neighboring fragments),
}
 and $\tns{z}$
 could be chosen to be unitary.
In particular, $\ket{\psi_{i_m}}$ might (theoretically) be a full configuration-interaction energy eigenstate of fragment $m$.
Overlaps between orbitals on different fragments would still require us to invoke the biorthogonal machinery, however.
Later, we will also suggest an approach in which $\tns{z}$ is not unitary,
 opening the door to less computationally expensive parameterizations.
The states described by $\tns{z}$ should also not necessarily be fragment eigenstates,
 but rather those that most efficiently and accurately describe interfragment correlations. 
Independent of the level of theory used for the fragment states, however, only a small fraction of the elements of
 $\tns{z}$ and $\bar{\tns{z}}$ will ever actually be computed in practice (even implicitly), due to
 truncations of the fragment spaces.

        \subsection{Single-fragment fluctuation operators}

We now construct explicit expressions for fluctuation operators that effect
 transitions of single fragments,
 regardless of the states of the other sub-systems.
\new{
 These can be thought of as excitations, de-excitations, {\etc}
}
We require that the action of $\opr{\tau}_{i_m}^{j_m}$ onto super-system basis state $\ket{\Psi_K}$ is as follows
 \begin{eqnarray}\label{action_definition}
  \opr{\tau}_{i_m}^{j_m}\ket{\psi_{k_1}\cdots\psi_{k_m}\cdots\psi_{k_N}}
                                        = \delta_{j_m,k_m}\ket{\psi_{k_1}\cdots\psi_{i_m}\cdots\psi_{k_N}}
 \end{eqnarray}
This action is reminiscent of a number-conserving pair of field operators onto a single-determinant electronic state,
 such that the null state results if the upper (``destruction'') index corresponds to an ``empty'' fragment state.
As shown, the lower and upper indices must refer to (potentially identical) states of the same fragment.

Operators obeying the above requirement will have the following commutation relation by definition
 \begin{eqnarray}\label{the_commutator}
  \big[\opr{\tau}^j_i,\opr{\tau}^l_k\big]_-
   = ~ \delta_{jk}\,\opr{\tau}^l_i ~-~ \delta_{il}\,\opr{\tau}^j_k
 \end{eqnarray}
This is shown from \eqnref{action_definition}
 by noting, first, that operators on different fragments commute,
 and, second, that a string of two operators on the same fragment gives the null state
 if the upper index of the left operator does not match the lower index of the right operator.
This is an important property concerning manipulation of fluctuation operators with techniques analogous to those familiar from 
 field operators.
We make use of this in the companion paper, which presents an excitonic coupled-cluster method.


It may seem desirable to make use of $\{\opr{\psi}_{i_m}\}$ and $\{\opr{\psi}^{i_m}\}$ directly
 to define state-to-state transition operators.
However, these operators do not obey the requisite canonical (anti)commutation relationships with their conjugates
 (consider especially the case of operators for differing particle numbers).
Such relationships would be necessary to generalize the notion of creation and annihilation operators \cite{Anderson:1994:CanonicalTrans},
 and these operators consequently do not transform as field operators with respect to changes of the many-electron basis.
Although these apparent many-electron creation operators will be convenient later,
 reformulation of the coming fluctuation-operator definitions in terms of them would
 necessarily retain a reference to the absolute vacuum dyadic $\ket{~}\bra{~}$.
Since we cannot abandon the language of state dyadics for the fluctuations,
 we proceed entirely in a state-space notation.

The following may then be regarded as a definition of a sub-system fluctuation operator on fragment $m$
 \begin{eqnarray}\label{fluctopcorr}
  \opr{\tau}_{i_m}^{j_m} = \sum_{k_1}\cdots\sum_{k_{m-1}}\sum_{k_{m+1}}\cdots\sum_{k_N} \,
                            \ket{\psi_{k_1}\cdots\psi_{k_{m-1}}\psi_{i_m}\psi_{k_{m+1}}\cdots\psi_{k_N}}
                            \bra{\psi^{k_1}\cdots\psi^{k_{m-1}}\psi^{j_m}\psi^{k_{m+1}}\cdots\psi^{k_N}} \n
 \end{eqnarray}
On account of the biorthogonality of the bases $\{\ket{\Psi_I}\}$ and $\{\ket{\Psi^I}\}$, a basis state acted upon by this operator
 will have non-zero projection onto, at most, one bra in the summation, and that will only happen if fragment $m$ is in state $\ket{\psi_{j_m}}$,
 which would then give unit coefficient to the super-system basis state that simply has $\ket{\psi_{j_m}}$ replaced by $\ket{\psi_{i_m}}$,
 as required by \eqnref{action_definition}.
Clearly, the choice of basis states for the single-fragment Fock spaces does not change this discussion.
We may therefore introduce an analogous set of operators defined with respect to the determinant
 bases, denoted for convenience as the set $\{\opr{\sigma}_P^Q\}$
 \begin{eqnarray}\label{fluctopdet}
  \opr{\sigma}_{P_m}^{Q_m} = \sum_{R_1}\cdots\sum_{R_{m-1}}\sum_{R_{m+1}}\cdots\sum_{R_N} \,
                            \ket{\phi_{R_1}\cdots\phi_{R_{m-1}}\phi_{P_m}\phi_{R_{m+1}}\cdots\phi_{R_N}}
                            \bra{\phi^{R_1}\cdots\phi^{R_{m-1}}\phi^{Q_m}\phi^{R_{m+1}}\cdots\phi^{R_N}} \n
 \end{eqnarray}


In \appref{completeness}, it is shown that either $\{\opr{\tau}_i^j\}$ or $\{\opr{\sigma}_P^Q\}$
 can be used to build a complete basis for the space of all super-system Fock-space operators.
In addition to abstractly assuring us that a Hamiltonian expansion in terms of fragment fluctuations is possible,
 any member of $\{\opr{\tau}_i^j\}$ must itself be resolvable in terms of $\{\opr{\sigma}_P^Q\}$, and vice versa.
Concretely, the transformation is seen to be rather simple.
Insertion
 of the resolutions the members of $\{\ket{\psi_i}\}$ and $\{\ket{\psi^i}\}$ in terms of the members of $\{\ket{\phi_P}\}$ and $\{\ket{\phi^P}\}$, or vice versa,
 into the definition of either $\opr{\tau}_{i_m}^{j_m}$ or $\opr{\sigma}_{P_m}^{Q_m}$ in \eqnsref{fluctopcorr}{fluctopdet}
 results in $N$$-$$1$ contractions of the diagonal blocks of $\tns{z}$ with $\bar{\tns{z}}$.
Resolving the consequent Kronecker deltas gives
 \begin{eqnarray}\label{preserves_rank}
  \opr{\tau}_{i_m}^{j_m}
                           &=& \sum_{P_m} \sum_{Q_m} ~ z^{P_m}_{i_m} \, \bar{z}^{j_m}_{Q_m} ~ \opr{\sigma}_{P_m}^{Q_m} \n
  \opr{\sigma}_{P_m}^{Q_m} &=& \sum_{i_m} \sum_{j_m} ~ \bar{z}^{i_m}_{P_m} \, z^{Q_m}_{j_m} ~ \opr{\tau}_{i_m}^{j_m}
 \end{eqnarray}
Effectively, each index transforms separately, with one power of $\tns{z}$ or $\bar{\tns{z}}$.
Inserting one of these identities into the other results in a truism.
Importantly, these transformations preserve fragment rank.

	\subsection{The Hamiltonian in terms of fragment fluctuation operators}

We may now resolve the Hamiltonian $\opr{\mathcal{H}}$ in terms of the single-fragment fluctuations.
For simplicity, we begin with the determinant basis.
%
%
%
%
In order to resolve the matrix elements $\bra{\Phi^P}\opr{\mathcal{H}}\ket{\Phi_Q}$,
 we first decompose the \textit{ab initio} expression for $\opr{\mathcal{H}}$
 in \eqnref{field_opH} as
 \begin{eqnarray}
  \opr{\mathcal{H}} = \opr{H}_1 + \opr{H}_2 + \opr{H}_3 + \opr{H}_4
 \end{eqnarray}
$\opr{H}_1$ collects together all terms from both the one-electron and two-electron parts of $\opr{\mathcal{H}}$
 that have all indices referring to orbitals of \textit{any} single fragment,
 and $\opr{H}_2$ similarly collects terms for all pairs of fragments (dimers).
It will also be useful 
 further decompose each of the $\opr{H}_M$ by separating out those
 terms that act on specific groups of fragments,
 for example, for $\opr{H}_4$,
 \begin{eqnarray}\label{further_decomposed}
  \opr{H}_4 = \sum_{m_1<m_2<m_3<m_4} \opr{H}^{(m_1,m_2,m_3,m_4)}
 \end{eqnarray}
 where $m_1$$<$$m_2$$<$$m_3$$<$$m_4$ under the summation runs over all unique tetramers.
The decomposition of $\opr{\mathcal{H}}$ truncates after $\opr{H}_4$, since there are a maximum of four orbital indices.
Notably, all terms in $\opr{H}_3$ and $\opr{H}_4$ must induce an inter-fragment charge transfer somewhere in the system.

\reworked{
In parallel to standard practice for matrix elements of one- and two-electron operators,
 it will be convenient to frame the discussion in terms of the number of
 fragments that have changed state in the bra, relative to the ket (henceforth, the number of \textit{substitutions}).
The logic for obtaining matrix elements closely mirrors the rules for determinant matrix elements of one- and two-electron operators.
For example, on account of the anticommutation rules for the field operators,
 we know that a matrix element of a given $\opr{H}_M$ will be zero if the number of
 substitutions is greater than $M$.
In addition to this, the fact that any term of any $\opr{H}_M$ operates on a maximum of two \textit{electrons} 
 places further restrictions on non-zero elements.
For example,
 $\opr{H}_4$ can have no non-zero matrix elements between states that differ by less than four fragment substitutions.
Similarly, $\opr{H}_3$ only has non-zero matrix elements between states that differ by either two or three
 substitutions
 (double substitutions represent a charge transfer in the average field of a third fragment).
}

\reworked{
To denote substitutions, the number of primes on a tuple index will be used to denote the number of substitutions relative to the unprimed index,
 and an overbar will denote a changed value of a sub-tuple therein.
For example, for two substitutions,
 $P'' = (P_1, \cdots \bar{P}_{m'} \cdots \bar{P}_{m''} \cdots P_N)$, where
 the unsubstituted tuple is
 $P = (P_1, \cdots P_{m'} \cdots P_{m''} \cdots P_N)$,
 and the fragments undergoing the substitution have been identified as $m'$ and $m''$.
We will always assume $m'$$<$$m''$$<$$\cdots$.
}
Recalling also that $\opr{\mathcal{H}}$ conserves overall particle number,
 we obtain the following expressions for the complete collection of non-zero matrix elements
 \begin{eqnarray}\label{nonzero_matelements}
   \bra{\Phi^P}\opr{H}_1\ket{\Phi_P} &=& \sum_m \bra{\phi^{P_m}} \opr{H}_1 \ket{\phi_{P_m}} \n
   \bra{\Phi^{P'}}\opr{H}_1\ket{\Phi_P} &=& \bra{\phi^{\bar{P}_{m'}}} \opr{H}_1 \ket{\phi_{P_{m'}}} \n
   \bra{\Phi^P}\opr{H}_2\ket{\Phi_P} &=& \sum_{m_1<m_2} \bra{\phi^{P_{m_1}}\phi^{P_{m_2}}} \opr{H}_2 \ket{\phi_{P_{m_1}}\phi_{P_{m_2}}} \n
   \bra{\Phi^{P'}}\opr{H}_2\ket{\Phi_P} &=& \sum_{m} \bra{\phi^{\bar{P}_{m'}}\phi^{P_m}} \opr{H}_2 \ket{\phi_{P_{m'}}\phi_{P_{m}}} \n
   \bra{\Phi^{P''}}\opr{H}_2\ket{\Phi_P} &=& \bra{\phi^{\bar{P}_{m'}}\phi^{\bar{P}_{m''}}} \opr{H}_2 \ket{\phi_{P_{m'}}\phi_{P_{m''}}} \n
   \bra{\Phi^{P''}}\opr{H}_3\ket{\Phi_P} &=& \sum_{m} (-1)^{\alpha^{P''}_m} \bra{\phi^{\bar{P}_{m'}}\phi^{\bar{P}_{m''}}\phi^{P_m}} \opr{H}_3 \ket{\phi_{P_{m'}}\phi_{P_{m''}}\phi_{P_m}} \n
   \bra{\Phi^{P'''}}\opr{H}_3\ket{\Phi_P} &=& \bra{\phi^{\bar{P}_{m'}}\phi^{\bar{P}_{m''}}\phi^{\bar{P}_{m'''}}} \opr{H}_3 \ket{\phi_{P_{m'}}\phi_{P_{m''}}\phi_{P_{m'''}}} \n
   \bra{\Phi^{P''''}}\opr{H}_4\ket{\Phi_P} &=& \bra{\phi^{\bar{P}_{m'}}\phi^{\bar{P}_{m''}}\phi^{\bar{P}_{m'''}}\phi^{\bar{P}_{m''''}}} \opr{H}_4 \ket{\phi_{P_{m'}}\phi_{P_{m''}}\phi_{P_{m'''}}\phi_{P_{m''''}}}
 \end{eqnarray}
\new{
 In the expansion of $\bra{\Phi^{P'}}\opr{H}_2\ket{\Phi_P}$, for example, $m'$ denotes the single fragment whose state is different in the bra and ket;
 since this fluctuation occurs in the average field of all other fragments, the states of those fragments (as they are given in $P$) are summed over.
}
These summations arise from the insertion of the decomposition in \eqnref{further_decomposed}, 
 after which, orbitals on ``spectator'' fragments contribute only factors of their unit biorthogonal overlaps.
Once the spectators are removed, the full $\opr{H}_M$ can be safely substituted for each of the terms of its prior decomposition according to \eqnref{further_decomposed},
 simply to declutter the notation.
For the sake of simplicity, the summations sometimes admit two copies of the same fragment
 state into a determinant; clearly, this evaluates to zero.
The summations also allow the states of the fragments to appear out of order inside of a bra;
 this does not contradict our established notation, which only insists that \textit{tuple}
 components are ordered.
Only in one case does the reordering lead to a sign change, and that is for a matrix element
 of $\opr{H}_3$, when the summation index $m$ is between $m'$ and $m''$ (since a charge must have been transferred);
 the exponent ${\alpha^{P''}_m}$ is one in this case and zero otherwise.

To complete the derivation, matrix elements of the form
 $\bra{\phi^{P_{m_1}}\cdots\phi^{P_{m_M}}}\opr{H}_M\ket{\phi_{Q_{m_1}}\cdots\phi_{Q_{m_M}}}$
 are multiplied
 with the products of $M$ single-fragment fluctuations $\{\opr{\sigma}_{P_m}^{Q_m}\}$ that effect the associated ``substitutions'' ($P_m$ and $Q_m$ might be equal).
Sums of such products can be used generate the same matrix elements
 as given in \eqnref{nonzero_matelements}, and these can therefore be used to build an exact representation of $\opr{\mathcal{H}}$.
Changing to the target basis of internally correlated states does not change the structure of this expression.
The transformation of the operators to the set $\{\opr{\tau}_{i_m}^{j_m}\}$, \textit{via} \eqnref{preserves_rank}, generates the
 contractions that define $\{\ket{\psi^{i_m}}\}$ and $\{\ket{\psi_{i_m}}\}$ in terms of $\{\ket{\phi^{P_m}}\}$ and $\{\ket{\phi_{P_m}}\}$ in \eqnsref{general_frag}{general_comp};
 we then finally arrive at
 \begin{eqnarray}
  \opr{\mathcal{H}} &=& \sum_m                 \sum_{\substack{I=(i_m)                             \\ J=(j_m)}}                          \; \bra{\Psi^I}\opr{H}_1\ket{\Psi_J} \, \opr{\tau}_{i_m}^{j_m} \n
                  &~& + \sum_{m_1<m_2}         \sum_{\substack{I=(i_{m_1},i_{m_2})                 \\ J=(j_{m_1},j_{m_2})}}                 \bra{\Psi^I}\opr{H}_2\ket{\Psi_J} \, \opr{\tau}_{i_{m_1}}^{j_{m_1}} \opr{\tau}_{i_{m_2}}^{j_{m_2}} \n
                  &~& + \sum_{m_1<m_2<m_3}     \sum_{\substack{I=(i_{m_1},i_{m_2},i_{m_3})         \\ J=(j_{m_1},j_{m_2},j_{m_3})}}         \bra{\Psi^I}\opr{H}_3\ket{\Psi_J} \, \opr{\tau}_{i_{m_1}}^{j_{m_1}} \opr{\tau}_{i_{m_2}}^{j_{m_2}} \opr{\tau}_{i_{m_3}}^{j_{m_3}} \n
                  &~& + \sum_{m_1<m_2<m_3<m_4} \sum_{\substack{I=(i_{m_1},i_{m_2},i_{m_3},i_{m_4}) \\ J=(j_{m_1},j_{m_2},j_{m_3},j_{m_4})}} \bra{\Psi^I}\opr{H}_4\ket{\Psi_J} \, \opr{\tau}_{i_{m_1}}^{j_{m_1}} \opr{\tau}_{i_{m_2}}^{j_{m_2}} \opr{\tau}_{i_{m_3}}^{j_{m_3}} \opr{\tau}_{i_{m_4}}^{j_{m_4}}
 \end{eqnarray}
The summations run over all unique monomers, dimers, trimers, and tetramers embedded within the overall super-system.
We have condensed this expression by using the super-system notation ($\ket{\Psi^I}$ and $\ket{\Psi_I}$) also for states of dimer sub-systems (\etc),
 including monomers.
It is interesting to note that this result can also be obtained 
 by starting directly with the fact that each of the $\opr{\psi}_{i_m}$ commutes with any term in $\opr{\mathcal{H}}$ that does not act on fragment $m$.
Though this is conceptually important, the ensuing detailed logic
 is essentially the same.
The numerical evaluation of the excitonic matrix elements now
 requires explicit
 insertion of terms from the \textit{ab initio} expansion of the Hamiltonian, but, conveniently, for small numbers of fragments.

As discussed in \appref{completeness}, since the basis of all possible fluctuation products is overcomplete, this expression for the Hamiltonian is not unique,
 but we conjecture that it is the most compact.
The formally quartic scaling of the number of matrix elements
 is simply a reflection of the size of the two-electron integrals tensor.
One advantage of decomposing $\opr{\mathcal{H}}$ into the $\opr{H}_M$ is
 that
 the two-fragment components here do not require subtraction of any double-counted single-fragment energies, {\etc}

		\section{Calculation and Approximation of Matrix Elements}

The remaining nontrivial matter
 is the practical evaluation of matrix elements that resolve excitonic Hamiltonians for real systems.
We will show that these matrix elements may be computed efficiently,
 using only one- and two-electron integrals 
 for small numbers of fragments ($\leq$$4$)
 and low-electron-order ($\leq$$2$), single-fragment data to represent the effects of intra-fragment correlation.
The single-fragment quantities are reusable and can be pre-computed in a single, linear-scaling step.
Further simplifications can be made for these coupling elements
 when fragments fall outside of overlap radius and eventually reach the asymptotic regime.
All of this is in spite of formally insisting on globally biorthogonalized bases.
The applicability of approximation schemes will also briefly be addressed.

The matrix elements we need are of the general form
 \begin{eqnarray}\label{x_mat_elem}
  \bra{\Psi^I} \opr{H}_M \ket{\Psi_J} = \bra{\psi^{i_{m_1}}\cdots\psi^{i_{m_M}}}\opr{H}^{(m_1,\cdots m_M)}\ket{\psi_{j_{m_1}}\cdots\psi_{j_{m_M}}}
 \end{eqnarray}
 for $M$ up to 4.
We recall that $\opr{H}^{(m_1,\cdots m_M)}$ 
 collects all terms of the \textit{ab initio} expression for $\opr{\mathcal{H}}$,
 where at least one field operator references an orbital on each of the 
 fragments $m_1$ through $m_M$, and no other fragments.
This can be written as a generic term-by-term expansion
 \begin{eqnarray}\label{decompH}
  \opr{H}^{(m_1,\cdots m_M)} = \sum_{l} h_l^{(m_1,\cdots m_M)} \, \opr{b}_l^{(\underline{m_1},\cdots m_M)}\cdots\opr{b}_l^{(m_1,\cdots \underline{m_M})}
 \end{eqnarray}
 where $l$ simply enumerates the terms.
The operator $\opr{b}_l^{(\underline{m_1},\cdots m_M)}$ collects together those field operators of the $l$-th term of $\opr{H}^{(m_1,\cdots m_M)}$
 that reference orbitals belonging to the fragment $m_1$, whose index bears the underline (and so forth, for the other participating fragments).
To within a phase of $\pm{1}$ (to account for permutations of field operators), $h_l^{(m_1,\cdots m_M)}$ denotes the respective
 one- or two-electron integral from the \textit{ab initio} expansion of \eqnref{field_opH}.

A matrix element of the excitonic Hamiltonian
\new{
 from \eqnref{x_mat_elem}
}
 may then be decomposed as
 \begin{eqnarray}\label{buildHpdt}
  \bra{\psi^{i_{m_1}}\cdots\psi^{i_{m_M}}}\opr{H}^{(m_1,\cdots m_M)}\ket{\psi_{j_{m_1}}\cdots\psi_{j_{m_M}}} = ~~~~~~~~~~~~~~~~~~~~~~~~~~~~~~~~~~~~~~~~~~~~~~~~~~~ \n 
    \sum_{l} \, (-1)^{\beta_{B_l}^{I,J}} \, h_l^{(m_1,\cdots m_M)} \, 
      \bra{\psi^{i_{m_1}}} \opr{b}_l^{(\underline{m_1},\cdots m_M)} \ket{\psi_{j_{m_1}}}
      \cdots
      \bra{\psi^{i_{m_M}}} \opr{b}_l^{(m_1,\cdots \underline{m_M})} \ket{\psi_{j_{m_M}}}
 \end{eqnarray}
 where an explanation of the phase factor therein is immediately forthcoming.
To show that this is possible, we first consider generic matrix elements of the form
 \begin{eqnarray}\label{provefactorizes}
  \bra{\phi^{P_{m_1}}\cdots\phi^{P_{m_M}}} \opr{b}^{(m_1)} \cdots \opr{b}^{(m_M)} \ket{\phi_{Q_{m_1}}\cdots\phi_{Q_{m_M}}} = ~~~~~~~~~~~~~~~~~~~~~~~~~~~~~~~~~~~~~~~~~~~~~~~~~~~ \n
  (-1)^{\beta_B^{P,Q}} \big\langle \, \big[\opr{\phi}^{P_{m_1}} \, \opr{b}^{(m_1)} \, \opr{\phi}_{Q_{m_1}}\big] \cdots \big[\opr{\phi}^{P_{m_M}} \, \opr{b}^{(m_M)} \, \opr{\phi}_{Q_{m_M}}\big] \, \big\rangle
 \end{eqnarray}
 where $\opr{b}^{(m)}$ can denote any string of single-electron field operators 
 (from the set $\{\opr{a}^p\}\cup\{\opr{c}_p\}$)
 that pertain to fragment $m$.
All of the operators that need to be permuted to reach the arrangement shown on the right-hand side of \eqnref{provefactorizes} either commute or anticommute
\new{
 (see the definitions of $\opr{\phi}_{P_m}$ and $\opr{\phi}^{P_m}$ in \eqnsref{antisymmpdt}{detcomplement}).
}
The accumulated phase exponent $\beta_B^{P,Q}$ then depends
 on the number of electrons on each fragment in both the bra ($P$) and in the ket ($Q$),
 and on the number of field operators in each of the $\opr{b}^{(m)}$,
 which are collected into a dependency on the tuple $B$.
(Actually, it only matters whether these numbers are even or odd.)
Most importantly, using the anticommutation relationships, the resulting vacuum expectation value factorizes into numbers that can each be computed individually,
 as vacuum expectation values of the operators for each fragment alone.
\new{
(Consider expanding each factor $\opr{\phi}^{P_m} \, \opr{b}^{(m)} \, \opr{\phi}_{Q_m}$ as a sum of strings that are normal ordered with respect to the vacuum,
 such that only the constant part of each survives the expectation value.)
}
By expanding the bra and ket on the left-hand side of \eqnref{buildHpdt} in the determinant basis
\new{
 (see \eqnsref{general_frag}{general_comp}),
}
 manipulating each of the resulting terms after the insertion of \eqnref{decompH} as just described
\new{
(\ie{\eqnref{provefactorizes}}),
}
 and then contracting the factors again with the expansion coefficients for the correlated fragment states,
 the right-hand side of \eqnref{buildHpdt} is obtained.
The final step of this sequence does require that 
 fragment states have definite particle number,
\reworked{
 in order that the phase of \eqnref{provefactorizes} is the same for each $l$
}
\new{
 ($\beta_{B_l}^{I,J} = \beta_{B_l}^{P,Q}$),
}
\reworked{
 and thus factors out;
}
 however, it need not be the same number in the bra and the ket (in the case of matrix elements for charge-transfer fluctuations).

According to \eqnref{buildHpdt},
 all information about the fragment states in $\bra{\Psi^I} \opr{H}_M \ket{\Psi_J}$ is encapsulated in factors of the generic form
 \begin{eqnarray}\label{dens_mat_elem}
  \bra{\psi^{i_m}}\opr{b}^{(m)}\ket{\psi_{j_m}}
    &=& \sum_{P_m,Q_m} \bar{z}_{P_m}^{i_m} z_{j_m}^{Q_m} \big\langle \opr{\phi}^{P_m} \, \opr{b}^{(m)} \, \opr{\phi}_{Q_m} \big\rangle
 \end{eqnarray}
\new{
A crucial aspect of this is that, in spite of the general biorthogonal notation,
 the factor $\big\langle \opr{\phi}^{P_m} \, \opr{b}^{(m)} \, \opr{\phi}_{Q_m} \big\rangle$
 is simply a vacuum expectation value of a string of field operators that obey anticommutation relationships of
 precisely the same form as is familiar for orthonormal sets (see \eqnref{bo_anticomm}).
}
\reworked{
Although the (complement) orbitals involved are not localized to fragment $m$, their ``tails''
 on neighboring fragments to do not change the expectation value, relative to what one would obtain
 if the complements were computed in the absence of neighbors (for $m$ only),
 since the matrix of biorthogonal overlaps remains the identity.
Therefore the elements defined by \eqnref{dens_mat_elem} are properties of fragment $m$ alone, which we shall interpret shortly.
What does change with the inclusion of neighbors is not the value of the expression in \eqnref{dens_mat_elem},
 but rather the corresponding integrals tensors, with which these are contracted in \eqnref{buildHpdt}.
The entirety of the complexity of demanding a globally consistent set of monomer fluctuations while
 working with overlapping basis functions has been moved
 into transformations of the one- and two-electron integrals.
}
\new{
This separation of information about the internal electronic structure of fragments and their energetic interactions
 is a fundamental feature of the chosen biorthogonal complements to the model-space,
 whose peculiar suitability we foreshadowed in \secref{exchange_concept}.
}

\reworked{
The quantities defined by \eqnref{dens_mat_elem} can be considered as elements of generalized reduced transition-density matrices (\ie{tensors}) for the individual fragments.
They effectively trace out much of the detail of intra-fragment correlation,
 such that the computational cost of subsequent steps will not depend directly on the correlation models
 internal to the fragments (the contraction with the integrals will depend on the basis size, however).
Conventional transition-density matrices \cite{Davidson:1976:DensityMatrices} involve one creation and one annihilation operator.
Here, there are eight kinds of tensors, resulting from different kinds of sub-strings of the one- and two-electron operators.
Each transition-density tensor needs to be evaluated between every pair of fragment states whose difference in particle number matches that of the tensor type.
In the case of only one field operator, there is an interesting connection to Dyson orbitals \cite{FetterWalecka:2003:GFBible,Ortiz:2004:BruecknerDyson}.
The most computationally expensive tensor will be the transition-density analogue of the two-electron reduced density matrix.
Since these are single-fragment properties, these tensors may be pre-computed in
 a formally linear-scaling step (and the most expensive tensors, for four field operators,
 need not be stored, since they do not contribute to couplings between fragments).
}

\reworked{
Though the complexity of dealing with antisymmetry in an overlapping basis has now been isolated
 to using a biorthogonal representation of the one- and two-electron integrals,
 it should also be pointed out that it is also not strictly necessary to ever compute the global biorthogonalized complement orbital basis.
Since the real-space resolutions of the kinetic and Coulombic operators are local,
 only orbitals in close proximity to a group under consideration need to be projected out of each complement.}
\new{
Therefore, the computation of any given matrix element of the form given in $\eqnref{x_mat_elem}$ will involve only information
 that is local to the vicinity of the fragments involved.
}

\new{
Given that the renormalized matrix elements may all be computed in independent calculations (making the implementation of these computations also easily parallelizable),
 the only relevant question as to the scaling of this step is the number of elements to be computed.}
With the connection to the primitive one- and two-electron integrals elucidated, we may
 apply a familiar analysis to determine the computational complexity of obtaining the excitonic Hamiltonian.
The matrix elements for spatially localized fragments will inherit the coarse features of the \textit{ab initio} Hamiltonian in a local basis.
For example, for the tetramer terms, which must involve two disjoint charge transfers,
 each acceptor fragment must be in the immediate vicinity of a donor fragment;
 however, the two donors may be quite far from each other before the overall interaction is negligible.
A thorough such analysis reveals that
 there are a quadratically scaling number of non-negligible terms in the excitonic Hamiltonian at the mesoscopic scale (linear in the bulk limit),
 for any fixed error tolerance.
This quadratic scaling of the interaction kernel is consistent with expectations from classical molecular mechanics.
Furthermore, as with the primitive two-electron integrals, all but a linear-scaling number of matrix elements may be cast as an
 electrostatic interaction between charge distributions that are outside of overlap (exchange) range of each other.
These can be expressed in terms of one-electron transition densities,
 which involve contractions of (multi-fragment) one-electron transition-density tensors with orbital-product distributions.
Such densities are importantly amenable to multipole approximations in the far field.
Although constructing such densities will sometimes involve orbitals on multiple fragments,
 there are nevertheless a linear-scaling number of them,
 due to the exponential decay of these densities with distance.
\new{
It is important to distinguish these statements from claims about associated wavefunctions, which can be arbitrarily complex
 in a local representation, depending on the relative magnitudes of fragment-energy gaps and inter-fragment couplings.
Making excitonic Hamiltonians applicable to conductors or frustrated systems (if possible) would require additional layers of formalism.
}

We finally turn our attention to the internally correlated fragment states themselves.
With respect to approximation schemes,
 it suffices to simply recall that the matrices $\tns{z}^{(m)}$ for each fragment
 were only ever required to be invertible.
Any existing method in electronic structure theory that can accomplish this
 and supply the necessary transition-density tensors
 can be applied (and the fragments are not required to all use the same method).
Notably, this opens the door for using the robust and efficient 
 equation-of-motion coupled-cluster theory \cite{Koch:1990:LRCC,Stanton:1993:EOMCC} for the fragment states.
In practice, the intention is to only ever (implicitly or explicitly) compute 
 a small number of columns and rows of each $\tns{z}^{(m)}$ and $\bar{\tns{z}}^{(m)}$, respectively.

		\section{Conclusion}

We have shown that it is possible to exactly write the
 electronic Hamiltonian for a super-system in terms of fluctuations of fragments between 
 internally correlated states,
 rigorously accounting for inter-fragment electron exchange and charge transfer 
 (also with a linearly dependent orbital basis).
This has the potential to fold the vast majority of the complexity of a wide variety of electronic
 structure problems into the low-scaling step of obtaining an effective Hamiltonian,
\new{
 particularly for non-covalently interacting systems.
}
The full range of familiar ground- and excited-state electronic structure methods, and their associated approximations,
 are readily applied to this Hamiltonian.

Explicit recipes have been given for constructing all necessary matrix elements,
 of which there are only an asymptotically quadratically scaling number, for a given threshold.
Constructing the Hamiltonian requires independent calculations on small groups of fragments ($\leq$$4$),
 using one- and two-electron integrals that are transformed to reflect the presence of other fragments.
All of these calculations may be cast 
 in terms of an overall linear-scaling number of reusable single-fragment tensors.
Furthermore, all but a linear-scaling number of matrix elements may be efficiently decomposed in terms of electrostatic
 interactions between a linear-scaling number of transition densities,
 which are further amenable to multipole approximations in the far field.

There are many features of this framework that hold promise for
 building
 finely tunable and efficient methods to capture properties of systems interacting
 with a large number of other systems.
First, it could potentially decouple the quality of the one-electron basis from the cost of
 the global calculation, to the extent that the necessitated qualities and numbers of fragment states
 may be separate considerations.
Second, since the super-system framework
 is theoretically independent of the level of approximation used for the fragments,
 it is not inherently subject to the shortcomings of any given electron-correlation model.
It is therefore immediately applicable to systems where current methods perform well for the isolated fragments.
This should then provide a robust approach for incorporating difficult electronic structure problems into calculations
 on large systems, perhaps using multi-reference methods for small molecules undergoing reactions
 and less expensive methods for resolving the states of peripheral molecules,
 and eventually giving way to force-fields or an embedding potential.
Third, pressing further into unknown territory, it may be possible to parameterize or interpolate
 the excitonic Hamiltonian as a function of nuclear coordinates,
 or even generalize the framework to handle vibronic states of the fragments.

The Fock-space formulation for the fragments here allows for inter-fragment charge resonance.
This could even create covalent linkages between fragments, in principle.
In cases where charge resonance is clearly unimportant, the excitonic Hamiltonian can be immediately truncated 
 after dimer terms.
Since all fragment fluctuations
 are described by orbitals local to a fragment, it will be interesting to explore
 the characteristics of this method with respect to basis-set superposition error,
 or possibly the lack thereof.
Although concrete algorithms are presently envisioned to work in overall number-conserving spaces,
 the ability of the formalism to handle open systems could be advantageous in the future.

At a technical level, it is interesting that our approach to inter-fragment exchange 
 has forced us to work with fragment Fock spaces,
 even when the model space is chosen to nominally conserve fragment charges.
This requirement arises from the biorthogonalization of the one-electron bases,
 which implicitly introduces charge-transfer components into the super-system complement basis.
This is somewhat intuitive;
 the fundamental ambiguity in fragment location of an electron in a nonorthogonal basis
 is intimately related to the concept of charge transfer.

	\section*{Acknowledgements}

The authors gratefully acknowledge start-up support from the Hornage Fund at the University of the Pacific,
 as well as equipment and travel support provided by the Dean of the College of the Pacific.
Helpful conversations with Christopher D.~Goff and Evgeny Epifanovsky are gratefully acknowledged.

\clearpage
{\LARGE \textbf{Appendix}}
\appendix

	\section{Completeness of Fluctuation Operators \label{completeness}}

The action of any operator in a space is fully determined by the collection of its matrix elements
 in a linearly independent basis for that space,
 or, equivalently, taken with respect to biorthogonal bases.
Therefore, in order to show that an arbitrary $\opr{O}$ in the super-system Fock space
 may be fully represented in terms of the set of all fluctuation operators $\{\opr{\tau}_i^j\}$,
 we need only to obtain such an expression for an operator
 wherein each matrix element is an independent degree of freedom.
This is the same as asserting that we can use this set to construct an operator that has only a single,
 arbitrarily chosen non-zero matrix element.
Choosing the $IJ$-th element to be non-zero, we define the operator $\opr{o}_I^J$, a component of $\opr{O}$, such that
 \begin{eqnarray}\label{defprimop}
  \bra{\Psi^{K}}\opr{o}_I^J\ket{\Psi_{L}} = \delta_{IK}\delta_{JL} = \prod_m \delta_{i_m,k_m}\delta_{j_m,l_m}
 \end{eqnarray}
This is easily accomplished by setting
 \begin{eqnarray}\label{resolveprimop}
  \opr{o}_I^J = \ket{\Psi_I}\bra{\Psi^J} = \prod_m \opr{\tau}_{i_m}^{j_m}
 \end{eqnarray}
Since the particular choice of fragment basis plays no role in this discussion, this conclusion applies equally well
 to the set $\{\opr{\sigma}_P^Q\}$.

\Eqnref{resolveprimop} essentially provides a recipe for constructing an arbitrary operator, one matrix element at a time.
Although any given operator may be represented as such,
 it does not have a unique resolution as a linear combination of \textit{all} products of fluctuation operators.
The set of all products of $N$ fluctuations (one for each fragment) is a complete and linearly
 independent basis for the operator space,
 as just demonstrated.
Adding to this set all products of lengths less than $N$ therefore builds a linearly dependent set.
A product of less than $N$ fluctuations has multiple non-zero matrix elements,
 and this is an important point as pertains to choosing compact representations of operators.
We intuitively expect the excitonic Hamiltonian to
 couple only small numbers of fragments simultaneously, and thereby also generate many non-zero matrix elements,
 analogous to
 the
 relatively short strings of field operators in the \textit{ab initio} representation.

	\section{Handling Linear Dependencies \label{lin_alg}}

\reworked{
Since the framework proposed in this work depends critically on both the
 ability to assign one-electron functions to specific fragments and on the linear independence of that basis,
 it behooves us to address the
 linearly dependent case.
This is important, as it is certain to arise when using diffuse functions for large systems.
}

\reworked{
The general problem is that the ambiguity in fragment location of an electron is complete 
 if the orbital it is in can be constructed as a linear combination
 of orbitals solely on fragments other than the one to which it is formally assigned.
We must somehow ``remove'' this linear dependency
 while simultaneously preserving the local structure of the fragment framework.
However, the direct removal of (nearly) null states
 from a one-electron basis produces orbitals that cannot
 generally be assigned to specific atoms or fragments.
Yet, removal of individual fragment-local or atom-local functions to resolve a linear dependency
 can create artificial asymmetries.
With some abstraction, the derivation in the main text
 can be applied directly to the linearly dependent case without modifying the one-electron basis.
We present here an alternate framing of that derivation,
}
 which is at a level of abstraction that allows us to handle both the linearly independent and linearly dependent cases together.
This allows us to show the path necessary to obtain working expressions 
 for the linearly dependent case without explicitly rederiving all details.

For any set of many-electron states $\{\ket{\Psi_I}\}$, linear dependence notwithstanding,
 we may write a resolution of the identity in the following form
 \begin{eqnarray}\label{psiidentity}
  1 = \sum_I \ket{\Psi_I}\bra{\Psi^I}
 \end{eqnarray}
 for some (potentially non-unique) choice of the set $\{\ket{\Psi^I}\}$.
By inserting this resolution of the identity into the time-independent Schr\"{o}dinger equation solved by $\ket{\Psi_\text{eigen}}$, we arrive at
 the matrix eigenvalue equation
 \begin{eqnarray}
  \tns{H}\tnsgrk{\Psi}_\text{eigen} = E_\text{eigen} \tnsgrk{\Psi}_\text{eigen}
 \end{eqnarray}
 where $\tnsgrk{\Psi}_\text{eigen}$ is a column matrix with elements $\Psi^I_\text{eigen} = \prj{\Psi^I}\ket{\Psi_\text{eigen}}$,
 and the matrix $\tns{H}$ has elements $H^I_J = \bra{\Psi^I}\opr{H}\ket{\Psi_J}$.
In the linearly dependent case, there are clearly some redundant degrees of freedom, and multiple choices of $\tnsgrk{\Psi}_\text{eigen}$
 can be used to represent the physical state $\ket{\Psi_\text{eigen}}$;
 however, for a given choice of $\{\ket{\Psi^I}\}$, the matrix $\tns{H}$ is fixed.
Assuming non-zero $E_\text{eigen}$, the eigenvector of $\tns{H}$ with eigenvalue of $E_\text{eigen}$ is unique.
(This can be generalized in the case of degenerate states.)
The physically redundant degrees of freedom
 are determined by the necessity that the projection of $\tnsgrk{\Psi}_\text{eigen}$ into the null space of
 $\tns{H}$ is zero.
($\tns{H}$ also has spurious eigenvectors with eigenvalue zero and no physical norm, corresponding to its null space.
 If the physical state $\ket{\Psi_\text{eigen}}$ happened to have eigenvalue of zero, these spurious
  eigenvectors could mix with it.
 This could be easily remedied by a scalar Hamiltonian shift, but we are generally concerned here with low-energy bound states,
  which have manifestly negative energy eigenvalues.)

In the case where $\{\ket{\Psi_I}\}$ is linearly independent, it is straightforward to show that the members of
 the set $\{\ket{\Psi^I}\}$ are the biorthogonal components uniquely determined by
 \begin{eqnarray}
  \bra{\Psi^I} = \sum_J \bar{S}^{IJ} \bra{\Psi_J}
 \end{eqnarray}
 where the elements $\bar{S}^{IJ}$ belong to the matrix $\bar{\tns{S}}=\tns{S}^{-1}$,
 where $\tns{S}$ is the matrix of overlaps $S_{IJ} = \prj{\Psi_I}\ket{\Psi_J}$.
Recalling that both $\tns{S}$ and $\bar{\tns{S}}$ are self-adjoint matrices, 
 we then straightforwardly arrive at the expected result that $\tns{H} = \bar{\tns{S}}\tilde{\tns{H}}$,
 where $\tilde{\tns{H}}$ has matrix elements $\tilde{H}_{IJ} = \bra{\Psi_I}\opr{H}\ket{\Psi_J}$.
The remainder of the work (already completed)
 can then be viewed as an exercise in constructing explicit forms of the members of $\{\ket{\Psi^I}\}$
 in terms of the fragment basis and underlying orbitals, such that we may write $\opr{H}$ in terms of fragment fluctuations with explicit
 formulas for the necessary scalar coefficients (matrix elements).

In the linearly dependent case, we will find it convenient to define fluctuation operators directly in the auxiliary space of coefficients.
Let the set of column matrices $\{\tnsgrk{\Psi}_I\}$ represent the orthonormal basis vectors of this \textit{coefficient} space, such that
 \begin{eqnarray}
  \tnsgrk{\Psi}_\text{eigen} &=& \sum_I \tnsgrk{\Psi}_I \Psi^I_\text{eigen}\n
          &=& \sum_{i_1} \cdots \sum_{i_N} \tnsgrk{\Psi}_{(i_1,\cdots i_N)} \Psi^{(i_1,\cdots i_N)}_\text{eigen}
 \end{eqnarray}
 where, in the second line, we remind ourselves of the structure of the index $I$ as a tuple that is decomposable in terms of fragment-state labels.
Let us now define a set of fluctuation matrices $\{\tnsgrk{\tau}^j_i\}$ with the following action
 \begin{eqnarray}
  \tnsgrk{\tau}^{j_m}_{i_m} \tnsgrk{\Psi}_{(k_1,\cdots k_m, \cdots k_N)} = \delta_{j_m,k_m} \tnsgrk{\Psi}_{(k_1,\cdots i_m, \cdots k_N)}
 \end{eqnarray}
 where, as with the physical states, these fluctuation matrices may be defined as superpositions of appropriate dyads.
Similarly, the proof that any matrix in the coefficient space may be written as a sum of products of fluctuation
 matrices proceeds along the same lines as in \appref{completeness}, assuring us that $\tns{H}$ may be built from such fluctuations.
The precise fragment-wise structure of this expansion of $\tns{H}$ (and therefore $\tnsgrk{\Psi}_\text{eigen}$) again depends on determining the the elements
 $\bra{\Psi^I} \opr{H} \ket{\Psi_J}$.
The primary difference, relative to the linearly independent case, is that the choice of $\{\ket{\Psi^I}\}$ is not 
 unique, and the null space of $\tns{H}$ depends on this choice,
 consequently affecting its fluctuation-matrix expansion.

In spite of any linear dependencies, \eqnref{superbasis} still provides the definition of
 the antisymmetrized tensor-products of correlated fragment states in terms of the (potentially linearly dependent) set of orbital configurations.
Therefore, in the linearly dependent case, \eqnref{supercompl}
 is still a suitable \textit{choice} of the members of $\{\ket{\Psi^I}\}$ that satisfy \eqnref{psiidentity},
 in terms of a suitable \textit{choice} of $\{\ket{\Phi^P}\}$ that satisfy
 \begin{eqnarray}\label{phiidentity}
  1 = \sum_P \ket{\Phi_P}\bra{\Phi^P}
 \end{eqnarray}
 which is quickly verified by inserting these resolutions of $\ket{\Psi_I}$ and $\ket{\Psi^I}$ into \eqnref{psiidentity},
 and realizing that the invertibility of $\tns{Z}$ is independent of linear dependencies in the set of physical states.
The important consequence of this is that, once formulas for the elements of a matrix $\tns{H}'$ (with elements $H^{\prime P}_Q = \bra{\Phi^P}\opr{H}\ket{\Phi_Q}$)
 are known for a given choice of $\{\ket{\Phi^P}\}$, and their structure in terms of numbers of fragments coupled is analyzed,
 then the path to constructing $\tns{H}$ in terms of fluctuations $\{\tnsgrk{\tau}^j_i\}$ via application
 of fragment-local transformations is the same as presented in the main text for the linearly independent case.

As with the linearly independent case, the construction of explicit forms for the members of $\{\ket{\Phi^P}\}$
 can be straightforward, so long as the matrix $\tns{s}$, built from the one-electron overlaps $s_{pq} = \prj{\chi_p}\ket{\chi_q}$,
 can be partially inverted.
Let the matrix $\bar{\tns{s}}$ satisfy
 \begin{eqnarray}
  \tns{p}_1 &=& \tns{p}_1\bar{\tns{s}}\tns{s} = \tns{s}\bar{\tns{s}}\tns{p}_1 \n
  1 &=& \tns{p}_0 + \tns{p}_1
 \end{eqnarray}
 where $\tns{p}_0$ is the orthogonal projector onto the null space of $\tns{s}$, and $\tns{p}_1$ is the orthogonal projector onto the range (null space complement) of $\tns{s}$.
Notably, only the projection of $\bar{\tns{s}}$ onto the range of $\tns{s}$ is uniquely defined.
Because of this, $\bar{\tns{s}}$ is not necessarily self adjoint for any choice, but
 we do have $\tns{p}_1\bar{\tns{s}}\tns{p}_1 = (\tns{p}_1\bar{\tns{s}}\tns{p}_1)^\dagger$, which is  equally valuable.
For a given choice of $\bar{\tns{s}}$, a choice of the members of $\{\ket{\Phi^P}\}$ may be specified again by \eqnref{detcomplement},
 but using a linearly dependent set of orbitals that satisfy
 \begin{eqnarray}
  \bra{\chi^p} = \sum_q \bar{s}^{pq} \bra{\chi_q}
 \end{eqnarray}
 where $\bar{s}^{pq}$ is an element of $\bar{\tns{s}}$.

We may, as before, define $\opr{c}_p$ and $\opr{c}^p$ as the creation operators associated with $\ket{\chi_p}$ and $\ket{\chi^p}$, respectively, 
 and define $\opr{a}_p$ and $\opr{a}^p$ as their respective Hermitian conjugates.
The Hamiltonian again takes exactly the same form as in \eqnref{field_opH} in terms of these operators.
(This can be shown by elimination of transformations that express the linearly dependent orbitals,
 their complements, and all associated field operators, in terms of some hypothetical orthonormal basis.)
The difference is that there is now redundancy in the operator set, and the anticommutation rules do not follow.
A given string in the field-operator resolution of the Hamiltonian will act not only on the fragments to which the indices
 in that string belong, but also on any fragments whose orbitals can be linearly combined to build the orbitals in question.
If a given orbital is linearly independent of the rest of the basis (likely the majority), then only expected fragment is
 involved.
Regardless, this does not increase the fragment order of the Hamiltonian, but it increases the number of fragments that
 could be associated with a given orbital index in the \textit{ab initio} Hamiltonian, which is an intuitive consequence of having a linear dependency arise
 due to overlaps of diffuse orbitals on different fragments.

A final comment is worthwhile to connect back to the abstract framework of resolutions of the identity.
Similar to what we have done in the one-electron space, the matrices $\bar{\tns{S}}$ and $\bar{\tns{S}}'$ may be defined
 as satisfying
 \begin{eqnarray}
  \tns{P}_1\bar{\tns{S}}\tns{S} ~=~ \tns{S}\bar{\tns{S}}\tns{P}_1 &=& \tns{P}_1 ~=~ 1-\tns{P}_0 \n
  \tns{P}'_1\bar{\tns{S}}'\tns{S}' ~=~ \tns{S}'\bar{\tns{S}}'\tns{P}'_1 &=& \tns{P}'_1 ~=~ 1-\tns{P}'_0 
 \end{eqnarray}
 where $\tns{S}'$ has elements $S'_{PQ} = \prj{\Phi_P}\ket{\Phi_Q}$,
 and $\tns{P}_0$ and $\tns{P}'_0$ are the orthogonal projectors into the null spaces of $\tns{S}$ and $\tns{S}'$, respectively.
For a given choice of $\bar{\tns{S}}$ and $\bar{\tns{S}}'$, the members of the sets $\{\ket{\Psi^I}\}$ and $\{\ket{\Phi^P}\}$ satisfy
 \begin{eqnarray}
  \bra{\Psi^I} &=& \sum_J \bar{S}^{IJ} \bra{\Psi_J} \n
  \bra{\Phi^P} &=& \sum_Q \bar{S}^{\prime PQ} \bra{\Phi_Q}
 \end{eqnarray}
which coincides with the unique choice of biorthogonal complements in the linearly independent case, when $\bar{\tns{S}}=\tns{S}^{-1}$ and $\bar{\tns{S}}'=\tns{S}'^{-1}$.
In the linearly dependent case, the use of these definitions in resolutions of the identity [\eqnsref{psiidentity}{phiidentity}] corresponds to additions of, and projections onto, null vectors.
If we make the choice that the arbitrary part of $\bar{\tns{s}}$ is zero, or equivalently that the arbitrary part of $\bar{\tns{S}}'$ is zero, which is likely
 the most convenient and practical choice,
 and we furthermore insist on \eqnref{supercompl} for our choice of the set $\{\ket{\Psi^I}\}$,
 then this is equivalent to choosing
 \begin{eqnarray}
  \bar{\tns{S}} = \bar{\tns{Z}}\tns{P}'_1\bar{\tns{S}}'\tns{P}'_1\bar{\tns{Z}}^\dagger
 \end{eqnarray}
This is clearly a valid pseudo-inverse of $\tns{S}=\tns{Z}^\dagger\tns{S}'\tns{Z}$, but it
 has non-zero components in the arbitrary part, due the fact that some eigenvectors of $\tns{S}'$ with non-zero eigenvalue
 transform to vectors that lie partly in the null space of $\tns{S}$, and vice versa.
The result of this is that the Hamiltonian matrix is
 \begin{eqnarray}
  \tns{H} = \bar{\tns{Z}}\tns{H}'\tns{Z}
 \end{eqnarray}
 which is a similarity transformation of the Hamiltonian in the configuration basis, as expected, but where
 \begin{eqnarray}
  \tns{P}_0\tns{H} \neq 0
 \end{eqnarray}
 meaning that $\tns{H}$ does not have the same null space as $\tns{S}$.
It still has a null space of the same dimension, but it corresponds to a specific, non-intuitive (but implicit and innocuous)
 choice of representation of the state vectors in the basis of tensor products of correlated states.

\end{document}